\documentclass[a4paper,aps,10pt,twocolumn,superscriptaddress,nofootinbib,amsmath,amssymb,floatfix]{revtex4-1}
\def\JCAPstyle#1{}
\usepackage{enumitem}
\usepackage{graphicx}
\usepackage[utf8]{inputenc}
\usepackage[T1]{fontenc}
\usepackage{cmap}
\usepackage{xcolor}
\usepackage{graphicx}
\usepackage{dcolumn}
\usepackage{bm}
\usepackage{bbm}
\usepackage{amsfonts}
\usepackage{amsmath}
\usepackage{enumitem}
\usepackage{lipsum}
\usepackage[utf8]{inputenc}
\usepackage[bottom]{footmisc}
\usepackage{cmap}
\usepackage{braket}
\usepackage{xcolor}
\usepackage{fmtcount}
      \usepackage[colorlinks]{hyperref}
\hypersetup{linkcolor=red,%
citecolor=blue,%
urlcolor=cyan}
\usepackage{url} 

\usepackage{dblfnote}
\usepackage{float}
\usepackage{cleveref}
\usepackage{natbib}
\usepackage[english]{babel}
\usepackage{blindtext}
\usepackage{comment}

\begin{document}

\title{Influences of modified Chaplygin dark fluid around a black hole}

\author{S. Zare }
\email{szare@uva.es}
\affiliation{Departamento de F\'{\i}sica Te\'orica, At\'omica y Optica and Laboratory for Disruptive Interdisciplinary Science (LaDIS), Universidad de Valladolid, 47011 Valladolid, Spain}
%ORCID: \href{https://orcid.org/0000-0003-0748-3386}{0000-0003-0748-3386}}
\author{L.M. Nieto}
\email{luismiguel.nieto.calzada@uva.es}
\affiliation{Departamento de F\'{\i}sica Te\'orica, At\'omica y Optica and Laboratory for Disruptive Interdisciplinary Science (LaDIS), Universidad de Valladolid, 47011 Valladolid, Spain}
%ORCID: \href{https://orcid.org/0000-0002-2849-2647}{0000-0002-2849-2647}}
\author{F. Hosseinifar}%
\email{f.hoseinifar94@gmail.com}
\affiliation{Department   of   Physics, Faculty of Science,   University   of   Hradec   Kr\'{a}lov\'{e},  Rokitansk\'{e}ho 62, 500   03   Hradec   Kr\'{a}lov\'{e},   Czechia}
%ORCID: \href{https://orcid.org/0009-0003-7057-451X}{0009-0003-7057-451X}}
\author{X.-H. Feng}
\email{xhfeng@tju.edu.cn}
\affiliation{Center for Joint Quantum Studies and Department of Physics, School of Science, Tianjin University, Tianjin 300350, China}
%ORCID: \href{https://orcid.org/0000-0003-3486-7828}{0000-0003-3486-7828}}
\author{H.  Hassanabadi}
\email{hha1349@gmail.com}
\affiliation{Department   of   Physics, Faculty of Science,   University   of   Hradec   Kr\'{a}lov\'{e},  Rokitansk\'{e}ho 62, 500   03   Hradec   Kr\'{a}lov\'{e},   Czechia}
%ORCID: \href{https://orcid.org/0000-0001-7487-6898}{0000-0001-7487-6898}}

%\date{\today}

\begin{abstract}
In this work, we study a static, spherically charged AdS black hole within a modified cosmological Chaplygin gas (MCG), adhering to the calorific equation of state, as a unified dark fluid model of dark energy and dark matter. 
We explore the influence of model parameters on several characteristics of the MCG-motivated charged AdS black hole (MCG-AdSBH), including the geodesic structure and some astrophysical phenomena such as null trajectories, shadow silhouettes, light deflection angles, and the determination of greybody bounds. 
We then discuss how the model parameters affect the Hawking temperature, remnant radius, and evaporation process of the MCG-AdSBH. Quasinormal modes are also investigated using the eikonal approximation method. Constraints on the MCG-AdSBH parameters are derived from EHT observations of M87* and Sgr A*, suggesting that MCG-AdSBH could be strong candidates for astrophysical black hole.
\medskip

\end{abstract}
\maketitle

\section{Introduction}\label{sec1}
Astronomical observations show that our universe is currently undergoing accelerated expansion, and dark energy with negative pressure and positive energy density is attributed to it \cite{PerlmutterApJ1999,RiessAJ1998,GarnavichApJ1998}. 
Dark energy was the first option used to illustrate negative pressure \cite{WangApJ2000,BahcallSci1999}.
Kiselev deduced the initial solution of the static and spherically symmetric black hole (BH) including quintessential matter \cite{KiselevCQG2003,RatraPRD1988,CaldwellPRL1998,SamiPRD2003}.
Some models combining dark matter and dark energy can be considered as a good candidate to explain the dark components of the Universe, with Chaplygin gas (CG) and its modification (MCG) being a suitable model to illustrate the observed accelerated expansion of the Universe \cite{KamenshchikPLB2001-1,BilicPLB2002,BentoPRD2002,KamenshchikPLB2001-1}.
In \cite{Sengupta2023} the CG in the Hubble tension is investigated, in \cite{AbdullahPRD2022} the growth of cosmological perturbations is studied, and in \cite{LiEPJP2020-1} the thermodynamic quantities for a static charged and spherically symmetric BH surrounded by such a model gas are explored.
The BH is investigated where it is immersed in a Chaplygin-type cosmological dark fluid, considering an additional parameter that influences the energy density of the fluid in \cite{Li2024} to obtain the geodesic structure, shadow and optical appearance of said BH.
Furthermore, the MCG is used to explore the stability of the Einstein-Gauss-Bonnet BH surrounded by this kind of gas \cite{LiEPJP2022-2} and the BH surrounded by MCG in the Lovelock gravity theory \cite{LiAP2022}. 
In reference \cite{Sekhmani2023} the authors considered of the MCG as a single fluid model unifying dark energy and dark matter, and constructed a static and spherically charged BH solution in the framework of general relativity.
Pure CG or generalized CG is a perfect fluid that behaves as a pressureless fluid at an early stage and as a cosmological constant at a later stage and is a good candidate for BH research. Recently, in \cite{ShadowRTheta}, the authors investigated the shadow, emission rate and deflection angle for a generalized Chaplygin-Jacobi dark fluid.

In this work, after introducing the metric in Sec.~\ref{sec0}, in Sec.~\ref{sec2}, we find the Hawking temperature and, consequently, the radius of the event horizon of a BH.
Then, using the photon geodesic equation, we determine the motion of light near the BH and the radius of the shadow for this specific metric in Sec.~\ref{sec3}.
In addition, we calculate the energy emission rate as a function of frequency, graybody factor, and emission power in Sections~\ref{sec4} and~\ref{sec5}.
Using the Eikonal approximation, in Sec.~\ref{sec6} we obtain the quasi-normal modes, and in Sec.~\ref{sec7} we calculate the amount of angular deflection of the light reaching the observer. In Sec.~\ref{sec8} we determine the evaporation time of the BH, and finally in Sec.~\ref{sec9} we present our conclusions.

\section{A Brief Review to Field Equation}\label{sec0}
According to Ref. \cite{Sekhmani2023}, we consider the following action for a charged source with a MCG structure within the framework of general relativity as
\begin{equation}
\mathcal{I}=\frac{1}{16 \pi} \int \mathrm{d}^4 x \, \sqrt{-g}\left[\mathcal{R}+\frac{6}{\bar{\ell}^{2}}-\frac{1}{4} F_{\mu \nu} F^{\mu \nu}\right]+\mathcal{I}_M,
\label{1}
\end{equation}
where $\mathcal{R}$ represents the Ricci scalar,  $g$ denotes the determinant of the metric tensor $g_{\mu\nu}$, $\ell$ denotes the AdS length, $ \mathcal{I}_M$ represents the matter contribution arising from the MCG background,  $F_{\mu \nu}$ is the field strength of the electromagnetic field, and $A_{\mu}$ is the gauge potential. By varying the action  \eqref{1} we obtain the following field equations
\begin{equation}
%\label{G}
G_{\mu\nu}-\frac{3}{\ell^2}g_{\mu\nu}=T_{\mu\nu}^{\text{EM}}+T_{\mu\nu}^{\text{MCG}},\quad
\partial_\mu(\sqrt{-g}F^{\mu\nu}) =0.
%\label{2}
\end{equation}
The symbols $G_{\mu\nu}$, $T_{\mu\nu}^{\rm MCG}$, and $T_{\mu\nu}^{\rm EM}$ represent the Einstein tensor, the energy-momentum tensor for MCG, and the energy-momentum tensor for the electromagnetic field, respectively, where $T_{\mu\nu}^{EM} = 2F_{\mu\lambda}F_{\nu}^{\lambda}-\frac{1}{2}g_{\mu\nu}F^{\lambda\delta}F_{\lambda\delta}$.
Considering a static, spherically symmetric, four-dimensional spacetime with a radial dependence for the zero component of the gauge potential (while the other components are zero), a second-order differential equation for the gauge potential is derived using the energy-momentum tensor for the electromagnetic field and exploiting the given spacetime background. The solution to this differential equation yields a compact solution for a Coulomb potential.
Furthermore, using the MCG, we adopt the equation of state given by $P=A\rho-B\rho^{-\beta}$, where $A$, $B$ are positive parameters and the parameter $\beta$ lies within the range $[0,1]$. 
By imposing conditions on the components of the energy-momentum tensors and considering all the components of the field equations, we derive the following expressions for the density and the lapse function
\cite{Sekhmani2023}	
	\begin{equation}
		\rho(r)=\Bigg\{\frac{1}{1+A}\Bigg(B+\bigg(\frac{\gamma}{r^3}\bigg)^{(1+A)(1+\beta)}\Bigg)\Bigg\}^{\frac{1}{1+\beta}},
		\label{22}
	\end{equation}
and
	\begin{eqnarray}\nonumber
		f(r) &=& 1-\frac{2
			M}{r}+\frac{Q^2}{r^2}+\frac{r^2}{\bar{\ell}^2}\\ && -\frac{r^2}{3} \bigg(\frac{B}{A+1}\bigg)^{\frac{1}{\beta+1}} \,
		_2F_1[\alpha, \nu; \lambda; \xi],
		\label{fr}
	\end{eqnarray}
where
	\begin{align}
		\alpha&=-\frac{1}{\beta+1},
		\,\,\,\nonumber
		\nu=-\frac{1}{1+A +\beta( A+1)}, \\
		\lambda&=1+\nu, \,\,\hspace{0.5cm}
		\xi=-\frac{1}{B}\left(\frac{\gamma}{r^3}\right)^{(A+1) (\beta+1)}.\label{frparameters}
	\end{align}
The metric has the following form \cite{book}
\begin{eqnarray}\label{ds2}
ds^2=-f(r)dt^2+\frac{1}{f(r)}dr^2+r^2 \left(d \theta^2+\sin ^2 \theta d \phi^2\right),
\end{eqnarray}

\section{Hawking Temperature and Remnant Radius}\label{sec2}

\begin{figure}[htb]
	\includegraphics[width=0.75\linewidth]{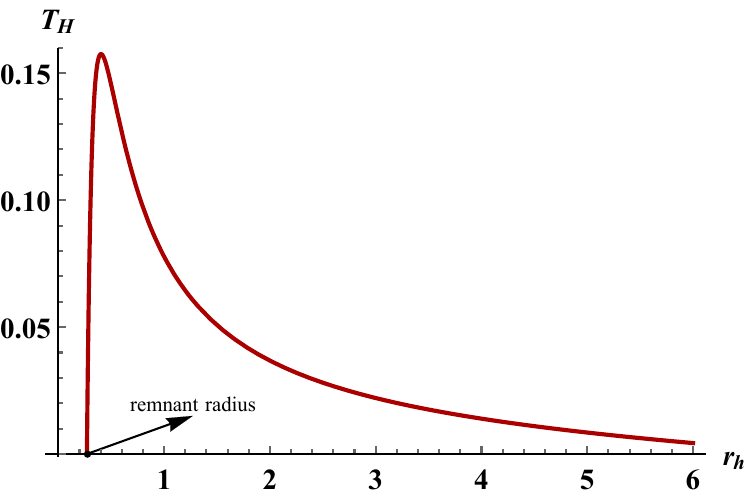}
	\caption{\label{fig:T_ER_8} Hawking temperature and remnant radius. The parameters are set as $A=1,\,B=0.1$, $\beta=0.1,\,\gamma=0.1,\,Q=0.01$ and $\bar{\ell} = 8$.} 
\end{figure}

The Hawking temperature near the event horizon of a BH with metric \eqref{ds2} is given by \cite{HawkingT}
\begin{eqnarray}
T_H=\frac{1}{4\pi}\frac{d}{dr}f(r)\Bigg|_{r=r_h}.
\end{eqnarray}
By setting $f(r)= 0$, the mass related to the horizon radius is calculated, so the temperature as a function of the horizon radius can be written as \cite{ZhangMT,Sekhmani2023}
\begin{align}\label{TH}
T_{\rm H}&=
2\frac{3 r_h}{8 \pi  \bar{\ell}^2}-\frac{Q^2}{4 \pi  r_{h}^3}+\frac{1}{4 \pi  r_{h}}\\
&
\ \ \ - \frac{r_{h}}{4 \pi } \left[\frac{A+1}{B}\right]^{\! -\frac{1}{\beta +1}} \left[ 1+\frac{\left(\frac{\gamma }{r_{h}^3}\right)^{(A+1) (\beta +1)}}{B}\right]^{\!-\frac{-1}{\beta +1}},
\nonumber
\end{align}
where $r_{h}$ is the horizon radius. The variation of the Hawking temperature as a function of $r_{h}$ is shown in Fig.~\ref{fig:T_ER_8}.
To calculate the remnant radius, we set the Hawking temperature obtained in Eq.~\eqref{TH} to zero $T_{\rm H}=0|_{r=r_{\rm rem}}$ \cite{AFilhoPLB2023}. In this case, we cannot find the explicit form of remnant radius with analytical methods. Substituting the remnant radius into the relation of $M$ in terms of $r_{\rm h}$, the remnant mass is obtained as
\begin{align}
\nonumber
M_{\rm rem} =& \frac{\dfrac{r_{\rm rem}^4}{\bar{\ell}^2}+Q^2+r_{\rm rem}^2}{2 r_{\rm rem}}\\
&-r_{\rm rem}^4 \left(\frac{B}{A+1}\right)^{\frac{1}{\beta +1}}\frac{ _2F_1\left(\alpha,\nu;\lambda;\xi_{\rm rem}\right)}{6 r_{\rm rem}}.
\end{align}

\section{Black hole shadow and constraints}\label{sec3}

\begin{figure}[htb]
\includegraphics[width=0.75\linewidth]{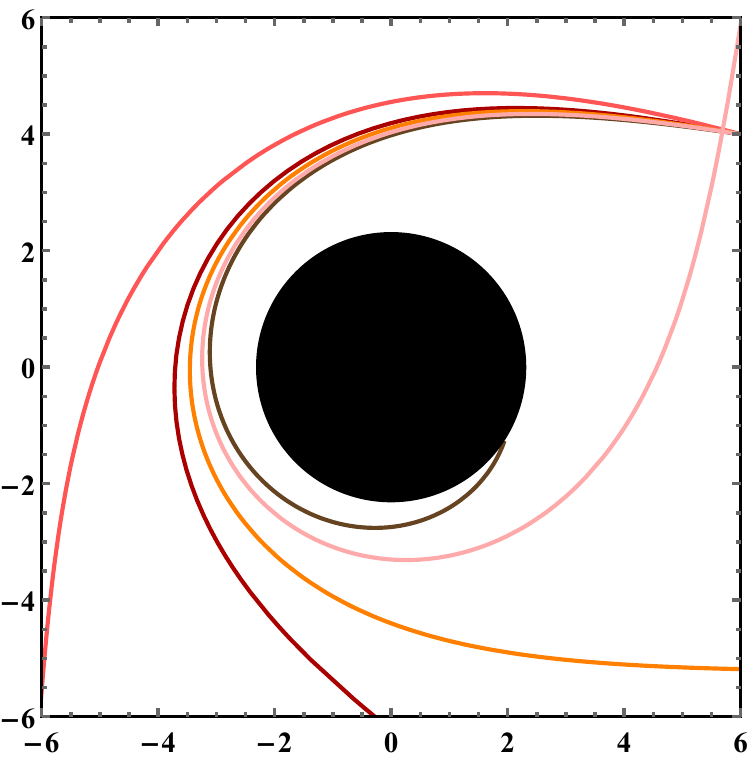}
\caption{\label{fig:traj} Light trajectory with variation of initial velocities. The parameters are chosen as $A=1$, $B=0.01$, $\beta=0.01$, $\gamma=0.01$, $Q=0.01, \,\bar{\ell}=6$ and $M=1$.} 
\end{figure}

\begin{figure}[htb]
	\includegraphics[width=0.75\linewidth]{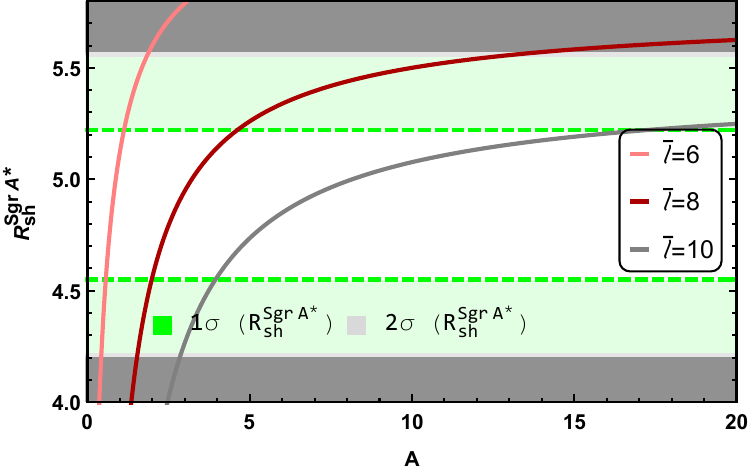}% Here is how to import EPS art
	\hfill
	\includegraphics[width=0.75\linewidth]{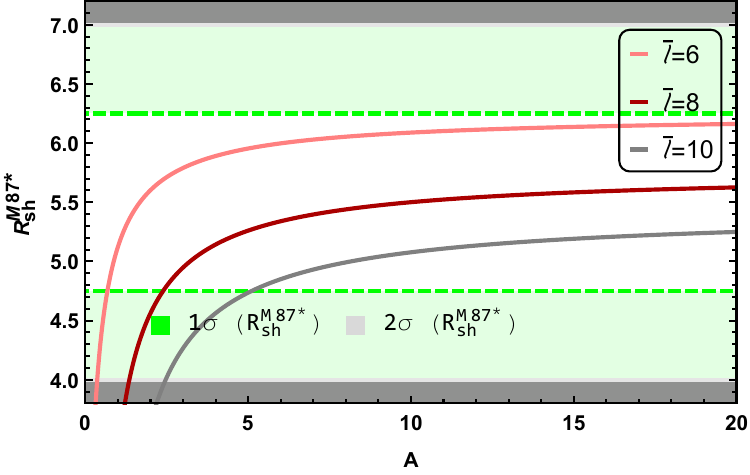}
	\caption{\label{fig:rshA} Shadow radius as a function of  $A$ for three different values of $\bar{\ell}$ with respect to the observations of Sgr $A^{*}$ and M87*, taking $B=0.1,\beta =0.1$, $\gamma =0.1,Q=0.01$ and $M=1$. The white and light green shaded regions align with the EHT images of Sgr A*  and M87* at $1\sigma$ and $2\sigma$ levels, respectively. }
\end{figure}

\begin{figure}[ht!]
	\includegraphics[width=0.75\linewidth]{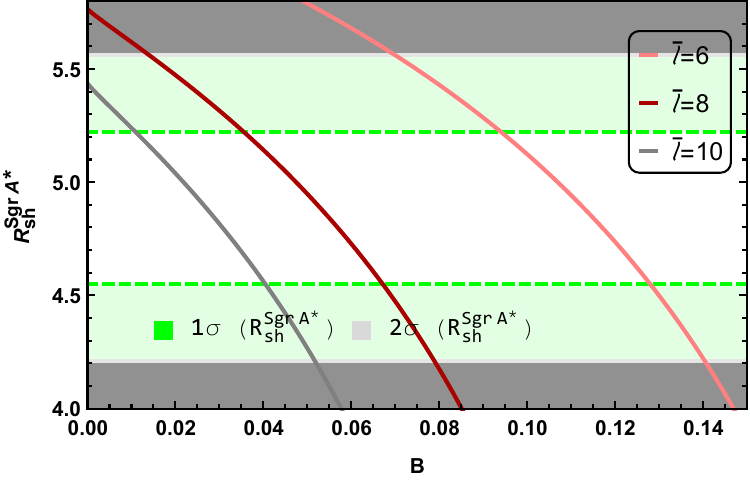}% Here is how to import EPS art
	\hfill
	\includegraphics[width=0.75\linewidth]{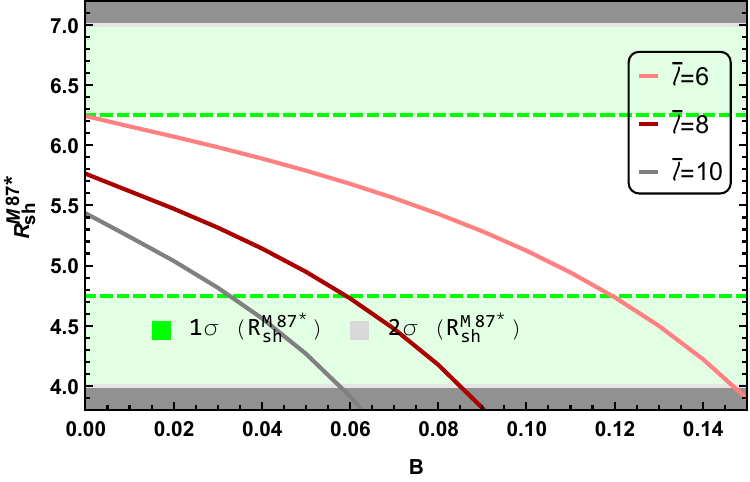}
	\caption{\label{fig:rshB} Shadow radius as a function of $B$ for three different values of $\bar{\ell}$ and for the Sgr A* and M87* observations, taking $A=1, \beta =0.1$, $\gamma =0.1, Q=0.01$ and $M=1$.}
\end{figure}

\begin{figure}[ht!]
	\includegraphics[width=0.75\linewidth]{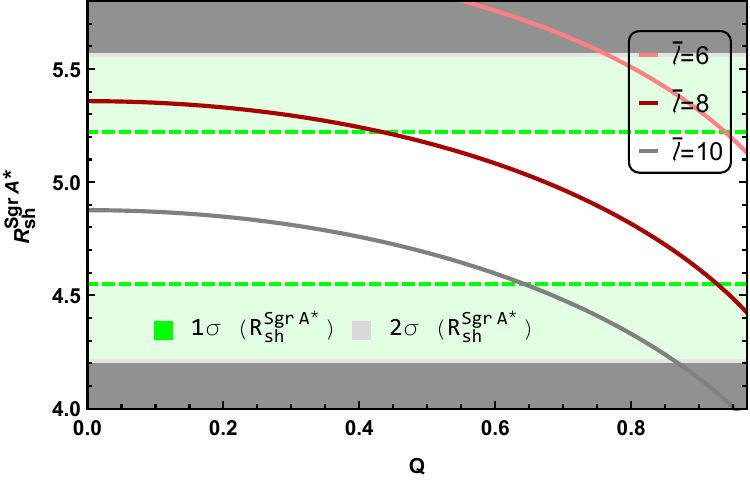}
	\hfill
	\includegraphics[width=0.75\linewidth]{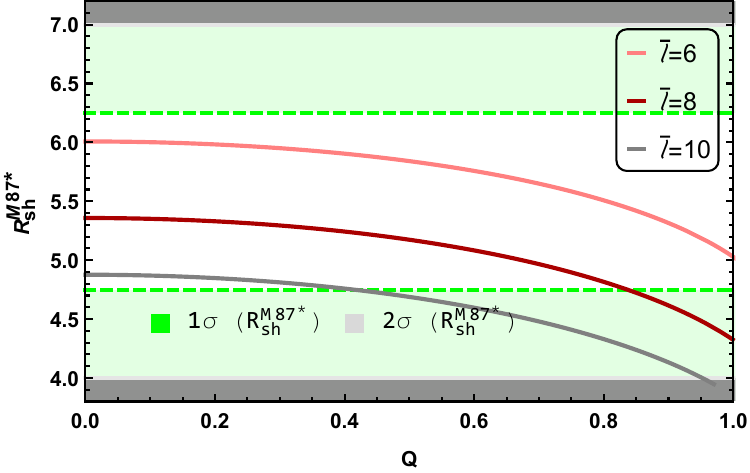} %0.494
\caption{\label{fig:rshQ}Change in shadow radius due to variation of $Q$ for  different values of $\bar{\ell}$, for Sgr A* and M87* observations, taking $A=5,B=0.1$, $\beta =0.05,\gamma =0.1,Q=0.01$, and $M=1$.} 
\end{figure}

\begin{figure}[htb]
	\includegraphics[width=0.75\linewidth]{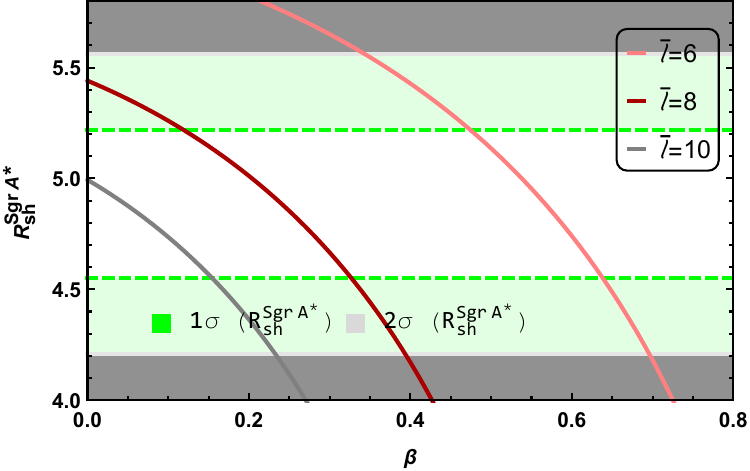}
	\hfill
	\includegraphics[width=0.75\linewidth]{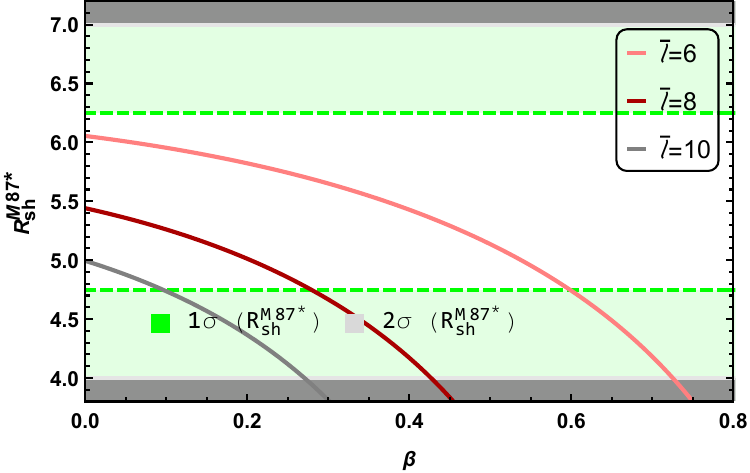}
	\caption{\label{fig:rshbeta}Shadow radius as a function of $\beta$ for three different values of $\bar{\ell}$ for the observations of Sgr A* and M87*, taking $A=5,B=0.1$, $\gamma =0.1,Q=0.01$ and $M=1$.} 
\end{figure}

To find the equation of motion for photons, the Euler–Lagrange equation $\frac{d}{d\tau}(\frac{\partial \mathcal{L}}{\partial \dot{x}^{\mu}})-\frac{\partial \mathcal{L}}{\partial x^{\mu}}=0$ is used, where  $\mathcal{L}=\frac{1}{2}g_{\mu\nu}\dot{x}^{\mu}\dot{x}^{\nu}$. Assuming $\theta = \frac{\pi}{2}$, the Lagrangian of the metric \eqref{ds2} becomes
\begin{eqnarray}
\mathcal{L} 
= \frac{1}{2}
\left(-f(r)\, \dot{t}^2+\frac{\dot{r}^2}{f(r)}+r^2\, \dot{\phi}^2\right)\label{L}.
\end{eqnarray}
There are two constants of motion $E$ and $L$ which are written in the form $E = f(r)\dot{t}$ and $L=r^2\dot{\phi}$. Therefore, it can be written as
\begin{eqnarray}
\dot{r}^2=E^2-L^2\frac{f(r)}{r^2}
\label{trajectory}.
\end{eqnarray}
Considering the quantity $ {L^2}  f(r)/r^{2}$ as the effective potential \cite{Li2024}, it can be expressed as
\begin{align}
\label{Veff}
V_{\rm Eff}=& \left(\delta+\frac{L^2}{r^2}\right)\left(1-\frac{2
   M}{r}+\frac{Q^2}{r^2}+\frac{r^2}{\ell^2}\right.\\ 
   &
\qquad\qquad\qquad -
  \left. \frac{r^2}{3} \bigg(\frac{B}{A+1}\bigg)^{\frac{1}{\beta+1}} \,
   _2F_1[\alpha, \nu; \lambda; \xi]\right),
\nonumber
\end{align}
where the parameter $\delta$ specifies the type of particle, which is either light-like ($\delta=0$) or time-like ($\delta=1$).  
\begin{figure}[ht!]
	\includegraphics[width=0.75\linewidth]{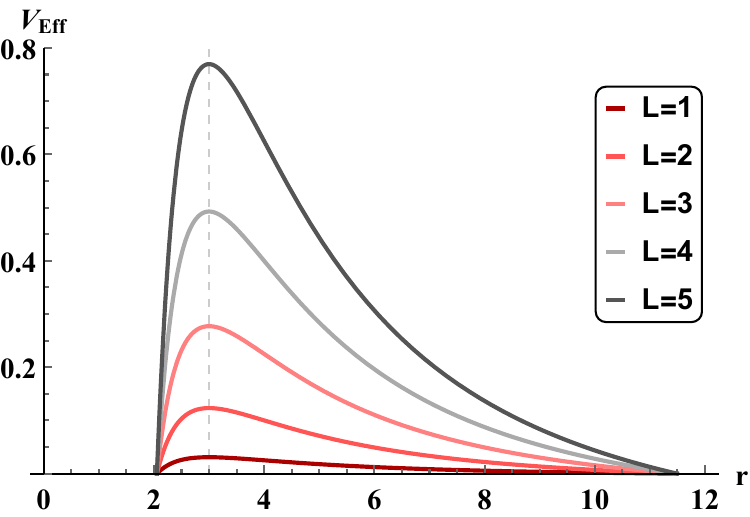}% Here is how to import EPS art
	\hfill
	\includegraphics[width=0.75\linewidth]{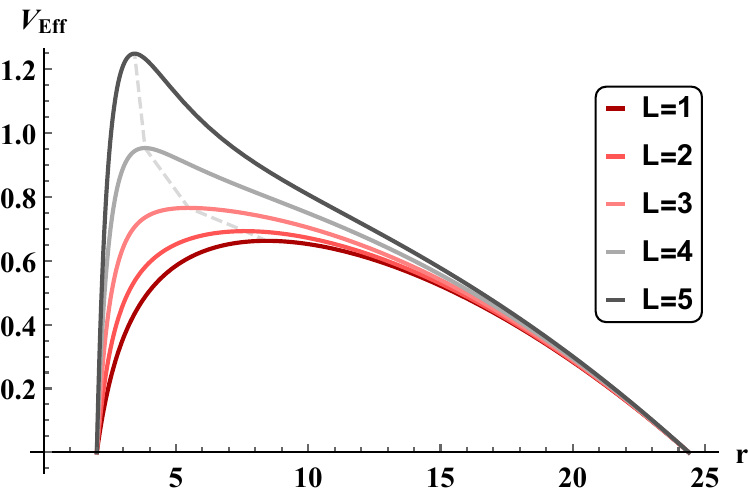}
	\caption{\label{fig:Veffr} Effective potential variation of light-like (first panel) and time-like (second panel) particles. The parameters have been chosen as  $A=1, \,B=0.1$,  $\beta=0.1, \,\gamma=0.1,$ $Q=0.01$, $\bar{\ell}=8$ and $M=1$.}
\end{figure}
Fig.~\ref{fig:Veffr} illustrates the effective potential versus radius.
In Fig.~\ref{fig:Veffr} we show the effective potential for light-like and time-like particles in the 
modified cosmological Chaplygin gas charged AdS black hole (MCG-AdSBH) background. The local maximum of the effective potential increases with higher values of angular momentum $L$. At this maximum, particles have unstable circular orbits. For time-like particles, as $L$ increases the maximum point appears at smaller values of $r$.
Using \eqref{trajectory} one can write the trajectory of light in the equatorial plane as
\begin{eqnarray}
\frac{d r}{d \phi} = \pm \sqrt{f(r)}\ \sqrt{\frac{r^2}{f(r)}\frac{E^2}{L^2}-1}\label{traj1} .
\end{eqnarray}
If $d r/d \phi =0$ is set, the minimum radius $r_{\rm min}$ would be found, which is assumed to be a turning point. Therefore, \eqref{traj1} can be rewritten in the form
\begin{eqnarray}
\frac{d r}{d \phi} = \pm \sqrt{f(r)}\ \sqrt{\frac{r^2}{f(r)}\frac{f(r_{\rm min})}{r_{\rm min}^2}-1}\label{traj}.
\end{eqnarray}
Using Eq.~\eqref{traj}, the light trajectory on the MCG-AdSBH is depicted in Fig.~\ref{fig:traj}.
Considering the circular orbit, one can write \cite{PerlickPR2022,CapozzielloJCAP2023,Capozziello}
\begin{eqnarray}
V_{\rm Eff}(r) \bigg|_{r=r_{\rm Ph}}=0=V'_{\rm Eff}(r)  \bigg|_{r=r_{\rm Ph}}.
\label{CircularOrbit}
\end{eqnarray}
Using \eqref{Veff} and~\eqref{CircularOrbit}, the radius of the photon sphere is calculated from \cite{RosaPRD2023}
$f'(r_{\rm Ph})r_{\rm Ph}^2-2r_{\rm Ph}f(r_{\rm Ph})=0$.
In this work, we have calculated the photon radius by numerical methods. Considering an observer that is located at the radius $r_{\rm o}$  of the BH, the shadow radius is obtained from~\cite{PanahEPJC2024}:
\begin{equation}
R_{\rm Sh}= \frac{r_{\rm Ph}}{\sqrt{f(r_{r_{\rm Ph}})}}\sqrt{f(r_{\rm o})}.
\end{equation}
We then derive constraints on the model parameters using observational Event Horizon Telescope (EHT)  data for M87* and Sgr A* based on their shadow images \cite{EHTL1,EHTL5,EHTL6,EHTL12,EHTL17,Kocherlakota,Vagnozzi,Zare}, as shown in Figs.~\ref{fig:rshA} to~\ref{fig:rshbeta}. 
Tables~\ref{Table:rshA} to~\ref{Table:rshbeta} present  lower and upper bounds on the parameters at $1\sigma$ and $2\sigma$ confidence levels.
%Table \ref{Table:rshA} Illustrates the limitation on parameter the $A$.
\begin{table}[ht!]
		\caption{The allowed range of  parameter $A$, based on observations of Sgr A* and M87* BH, for three different $\bar{\ell}$, taking $B=0.1,\,\beta =0.1,\,\gamma =0.1,\,Q=0.01$ and $M=1$.}
		\centering
		\label{Table:rshA}
		\begin{tabular}{|c|c|c|c|c|}
		\hline
		  & \multicolumn{2}{c|}{Sgr $A^{*}$} & \multicolumn{2}{c|}{M $87^{*}$}\\
		 \cline{2-5}
		 $A$ & $1\sigma$ & $2\sigma$ & $1\sigma$ & $2\sigma$\\
		 \cline{2-5}
		 & \multicolumn{4}{c|}{$(\text{lower bound, upper bound)}$}\\
		  \hline
		 $\bar{\ell}=6$ & $(0.52,\,1.13)$ & $(0.43,\,1.86)$ & $(0.53,\,-)$ &  $(0.36,\,-)$\\
		 \hline
		 $\bar{\ell}=8$ & $(1.83,\,4.61)$ & $(1.54,\,13.19)$ & $(2.39,\,-)$ & $(1.36,\,-)$\\
		 \hline
		 $\bar{\ell}=10$ & $(3.54,\,17.19)$ & $(2.87,\,-)$ & $(5.13,\,-)$ & $(2.46,\,-)$\\
		 \hline
		\end{tabular}
	\end{table}
\begin{table}[ht!]
		\caption{Allowed range of  $B$, based on information obtained from Sgr $A^{*}$ and $M 87^{*}$ BH, for three different $\bar{\ell}$, with $A=1$, $\beta =0.1,\gamma =0.1,Q=0.01$ and $M=1$.}
		\centering
		\label{Table:rshB}
		\begin{tabular}{|c|c|c|c|c|}
		\hline
		 & \multicolumn{2}{c|}{Sgr A*} & \multicolumn{2}{c|}{M87*}\\
		 \cline{2-5}
		 $B$ & $1\sigma$ & $2\sigma$ & $1\sigma$ & $2\sigma$\\
		 \cline{2-5}
		 & \multicolumn{4}{c|}{$(\text{lower bound, upper bound)}$}\\
		 \hline
		 $\bar{\ell}=6$ & $(0.09\,,0.13)$ & $(0.07,\,0.14)$ & $(-,\,0.12)$ & $(-,\,0.15)$\\
		 \hline
		 $\bar{\ell}=8$ & $(0.04,\,0.07)$ & $(0.01,\,0.08)$ & $(-,\,0.06)$ & $(-,\,0.09)$
		\\
		\hline
%		\hspace{0.2cm}
		 $\bar{\ell}=10$ & $(0.01,\,0.04)$ & $(-,\,0.05)$ & $(-,\,0.03)$ & $(-,\,0.06)$\\
		 \hline
		\end{tabular}
	\end{table}
\begin{table}[ht!]
		\caption{Allowed range of $Q$, based on documentation obtained  from observations of Sgr A* and M87* BH, for different $\bar{\ell}$, with $A=5,B=0.1,\beta =0.05,\gamma=0.1$ and $M=1$.}
		\centering
		\label{Table:rshQ}
		\begin{tabular}{|c|c|c|c|c|}
		\hline
		 & \multicolumn{2}{c|}{Sgr A*} & \multicolumn{2}{c|}{M87*}\\
		 \cline{2-5}
		 $Q$ & $1\sigma$ & $2\sigma$ & $1\sigma$ & $2\sigma$\\
		  \cline{2-5}
		  & \multicolumn{4}{c|}{$(\text{lower bound, upper bound)}$}\\
		 \hline
		 $\bar{\ell}=6$ & $(0.94,\,-)$ & $(0.77,\,-)$ & $(-,\,-)$ & $(-,\,-)$\\
		 \hline
		 $\bar{\ell}=8$ & $(0.43,\,0.96)$ & $(-,\,-)$ & $(-,\,0.84)$ & $(-,\,-)$
		\\
		\hline
%		\hspace{0.2cm}
		 $\bar{\ell}=10$ & $(-,\,0.72)$ & $(-,\,0.86)$ & $(-,\,0.41)$ & $(-,\,0.95)$\\
		 \hline
		\end{tabular}
	\end{table}
\begin{table}[ht!]
		\caption{Allowed range of $\beta$, based on documentation obtained  from observations of Sgr A* and M87* BH, for different $\bar{\ell}$, taking $A=5,B=0.1$, $\gamma =0.1,Q=0.01$, and $M=1$.} 
		\centering
		\label{Table:rshbeta}
		\begin{tabular}{|c|c|c|c|c|}
		\hline
		 & \multicolumn{2}{c|}{Sgr A*} & \multicolumn{2}{c|}{M87*}\\
		 \cline{2-5}
		 $\beta$ & $1\sigma$ & $2\sigma$ & $1\sigma$ & $2\sigma$\\
		 \cline{2-5}
		 & \multicolumn{4}{c|}{$(\text{lower bound, upper bound)}$}\\
		 \hline
		 $\bar{\ell}=6$ & $(0.47,\,0.66)$ & $(0.34,\,0.70)$ & $(0.12,\,0.35)$ & $(-,\,0.39)$\\
		 \hline
		 $\bar{\ell}=8$ & $(0.12,\,0.35)$ & $(-,\,0.39)$ & $(-,\,0.28)$ & $(-,\,0.42)$
		\\
		\hline
%		\hspace{0.2cm}
		 $\bar{\ell}=10$ & $(-,\,0.18)$ & $(-,\,0.23)$ & $(-,\,0.09)$ & $(-,\,0.27)$\\
		 \hline
		\end{tabular}
	\end{table}

\section{Energy emission rate}\label{sec4}

\begin{figure}[htb]
\includegraphics[width=0.75\linewidth]{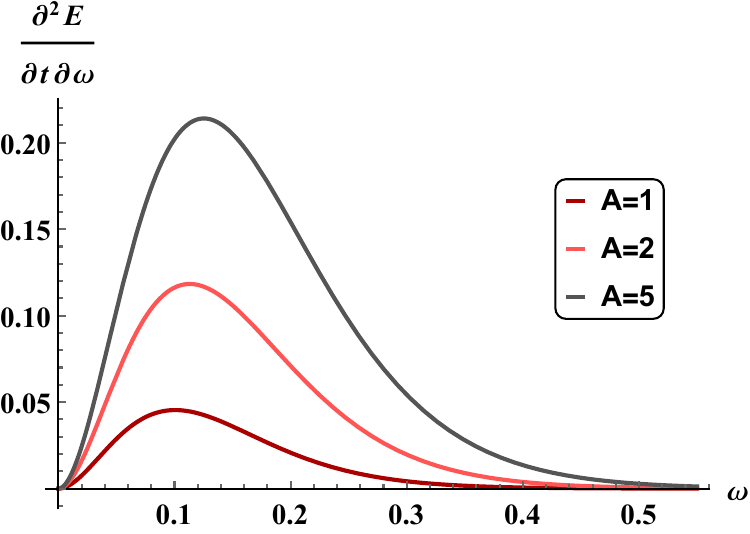}
\caption{\label{fig:ERnew_8} Plot of the energy emission rate as a function of $\omega$ with different $Q$s. The parameters are set as $A=1,\,B=0.1$, $\beta=0.1,\,\gamma=0.1,\,Q=0.01$ and $\bar{\ell} = 8$.} 
\end{figure}

The thermal radiation emitted by a BH is related to its Hawking temperature. Using the BH shadow, the energy emission rate per unit time and frequency is given by \cite{12}
\begin{eqnarray}
\frac{\partial^2 E}{\partial\omega\, \partial t} = \frac{2 \pi^2 \sigma \omega^3}{
 \exp(\omega/T_H) - 1},
\end{eqnarray}
where $\omega$ refers to the photon emission frequency, $\sigma$ represents the cross section and is equal to $\pi R_{\rm Sh}^2$, and $T_H$ shows the Hawking temperature calculated from Eq.~\eqref{TH}.  Fig.~\ref{fig:ERnew_8} shows the energy emission rate as a function of $\omega$.
Obviously, by increasing the parameter $A$, the maximum value of the emission power increases and the maximum occurs at a higher frequency.

\section{Greybody Bounds}\label{sec5}

It is possible for one of them to fall into the BH while the other escapes into space, resulting in radiation modified by the BH's gravitational field, known as the graybody effect, so the graybody factor measures the degree of this radiation deviation. A factor less than 1 indicates absorption or scattering within the BH. The graybody limit represents the maximum deviation of the radiation from the ideal blackbody spectrum. The lower limit of the graybody factor can be calculated as \cite{gbf2}
\begin{eqnarray}
T_l(\omega)\geq \text{sech}^2 \left(\frac{1}{2\omega}\int_{r_h}^{\infty}V(r) \frac{dr}{f(r)}\right).
\end{eqnarray}
The potential is given by \cite{gbf2}
\begin{eqnarray}
V(r)=f(r)\left(\frac{l(l+1)}{r^2}+\frac{1-s^2}{r}\frac{d}{dr}f(r)\right)\label{veffl},
\end{eqnarray}
where $l$ is the angular momentum and the parameter $s$ denotes the spin such that $s = 0$ gives the effective potential for the scalar perturbation and $s = 1$ expresses the potential for the electromagnetic perturbation.

\begin{figure}[htb]
	\includegraphics[width=0.75\linewidth]{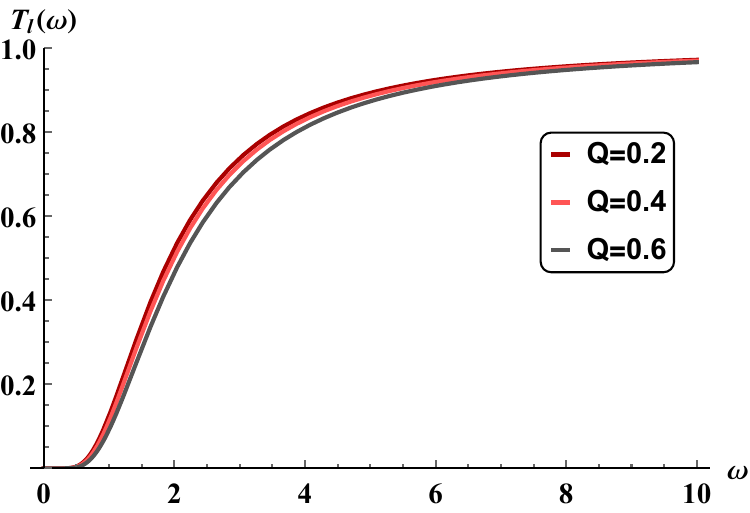}
	\hfill
	\includegraphics[width=0.83\linewidth]{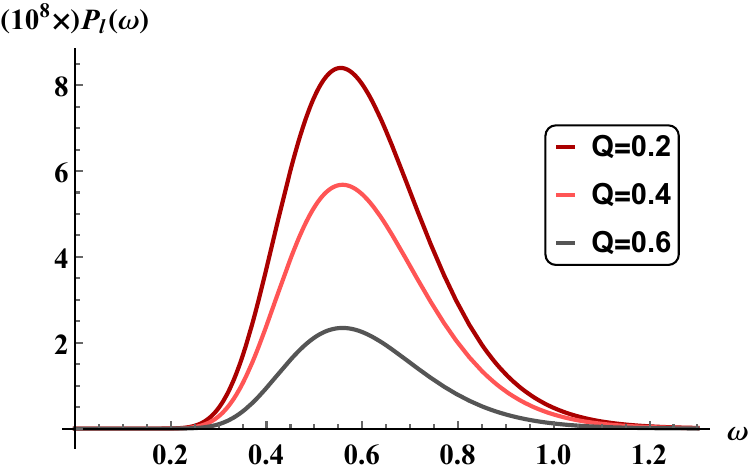}
	\hfill
	\includegraphics[width=0.75\linewidth]{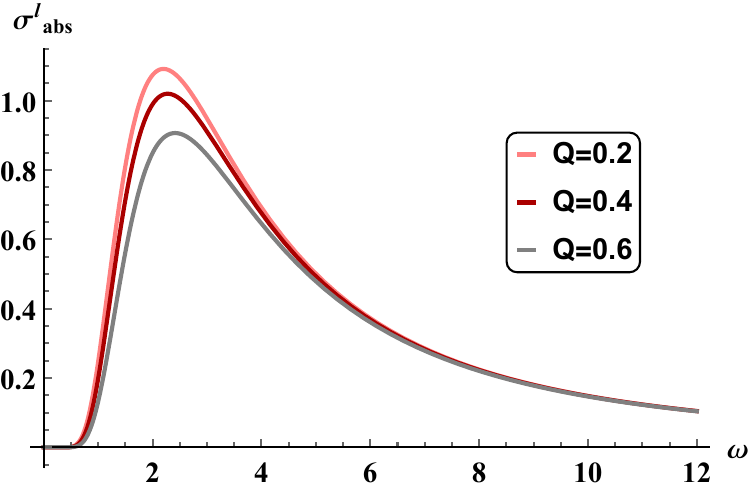}
	\caption{\label{fig:GBB} Plot of energy emission rate (first panel), emitted power (second panel) and cross section (third panel) as a function of $\omega$ with different $Q$. The parameters are set as $A=1,\,B=0.1$, $\beta=0.1,\,\gamma=0.1,\,Q=0.01$ and $\bar{\ell} = 8$.} 
\end{figure}

The power emitted by the $l$th mode is found as \cite{qnm2}
\begin{eqnarray}
P_{l}(\omega)=\frac{A}{8\pi^2}T_l(\omega)\frac{\omega^3}{\exp(\omega / T_{\rm H})-1},
\end{eqnarray}
where $A$ and $T_{\rm H}$ refer to the surface area of a sphere with radius $r_{\rm h}$ and Hawking temperature, respectively.
The illustration of the greybody bounds and its emitted power by electromagnetic perturbation are shown in the first and second panels of Fig.~\ref{fig:GBB}, respectively.
According to these graphs, as the parameter $Q$ increases, the transmission probability increases. Furthermore, the emitted power decreases as this parameter increases and its maximum occurs at a higher frequency. In this way, the partial absorption cross section is given by \cite{CrispinoPRD2009,AnacletoPLB2020}
\begin{eqnarray}
\sigma_{abs}^l =\frac{\pi (2l+1)}{\omega^2} |T_l(\omega)|^2,
\end{eqnarray}
which is shown in the third panel of Fig.~\ref{fig:GBB}. Increasing the parameter $Q$ reduces the maximum value of the absorption cross section and shifts this maximum to smaller values of $\omega$.

\section{Quasinormal Modes}\label{sec6}

\begin{figure}[htb]
	\includegraphics[width=0.75\linewidth]{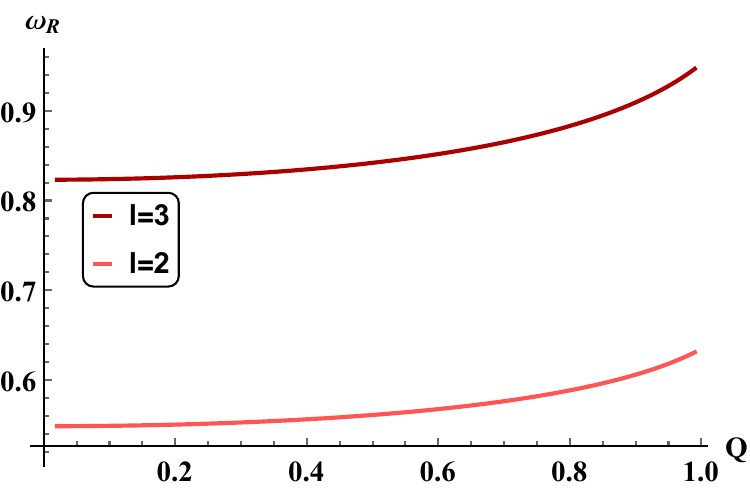}
	\hfill
	\includegraphics[width=0.75\linewidth]{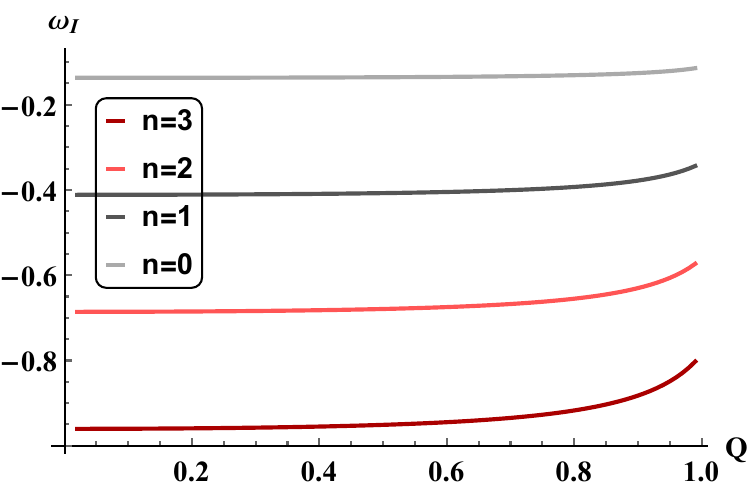}
	\caption{\label{fig:eikonalwQl1} Behavior of $\omega_R=\text{Re}(\omega)$ and $\omega_I=\text{Im}(\omega)$ as a function of $Q$. The other parameters are set as $A=1,\,B=0.01$, $\beta=0.01$, $\gamma=0.01,\,\bar{\ell}=5$ and $M=1$. $l=33$ for $\text{Im}(\omega)$.} 
\end{figure}

Applying perturbation on a BH causes damping oscillations called quasinormal modes. The frequency of these oscillations is a complex number whose real part shows the oscillation frequency and the imaginary part indicates the decay part that was applied by the perturbation \cite{qnm}. In the eikonal limit, $l\rightarrow\infty$, the potential of the equation~\eqref{veffl} reduces to \cite{eik2} $V_{0}^{E}=l^{2} f(r)/r^{2}$. The quasinormal frequency is denoted as
\begin{eqnarray}
\omega_n= \Omega\, l -i \Lambda  \left(n+\frac{1}{2}\right),
\end{eqnarray}
where $\Omega$ is the angular velocity and $\Lambda$ is the principal Lyapunov exponent, which are given by \cite{11,MouraEik}
\begin{align}
\nonumber \Omega&=\frac{\sqrt{f(r_c)}}{r_c},\\
\Lambda&=\sqrt{-\frac{r^2}{2}\left(f'(r)(\frac{f(r)}{r^2})'+f(r)(\frac{f(r)}{r^2})''\right)}_{r=r_c}.
\end{align}
The parameter $r_c$ is the maximum value of $V_0^{E}$. The behavior of $\omega$ versus the parameter $Q$ is obtained in Fig.~\ref{fig:eikonalwQl1}.

\section{Deflection Angle}\label{sec7}

\begin{figure}[htb]
\includegraphics[width=0.75\linewidth]{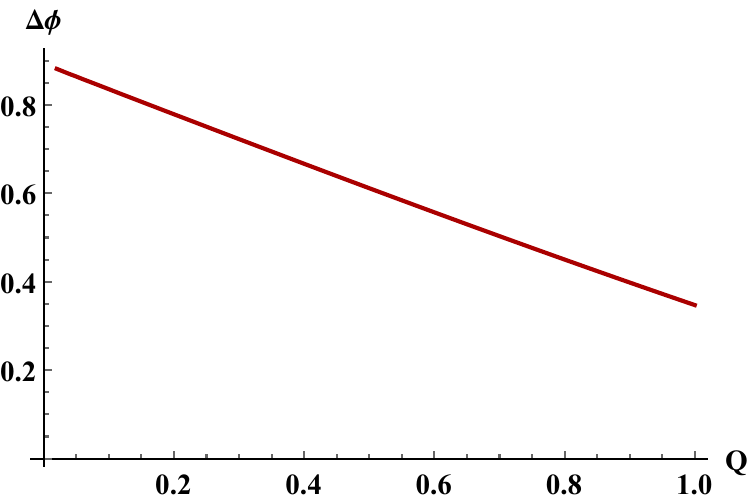}
\caption{\label{fig:defang} Variations of the deflection angle with $Q$.} 
\end{figure}

When light shines from a star and reaches an observer after passing through the vicinity of a BH, it undergoes gravitational lensing and gravitational deflection. The angle of deflection of this light, known as deflection angle, is given by \cite{DA12}
\begin{eqnarray}
\Delta \phi =2 \int_{r_{\min }}^{\infty } \frac{1}{r \sqrt{\frac{r^2}{b^2}-f(r)}} \, dr-\pi,
\end{eqnarray}
where $b$ referred to the impact parameter and is given by $r_{\rm min}/\sqrt{f(r_{\rm min})}$. The variation curve of deflection angle by setting parameters as $M=1,\, \bar{\ell}=8,\, A=1,\, B=0.1,\, \beta =0.1,\, \gamma =0.1$ and $b =10$, is illustrated in Fig.~\ref{fig:defang}.
As can be seen from Fig~\ref{fig:defang}, by increasing the parameter $Q$, the value of deflection angle decreases.

\section{Evaporation Process}\label{sec8}

The lifetime of a BH can be determined by integrating the change in its mass over time $\tau$, and is calculated as follows \cite{12,evap2}
\begin{eqnarray}\label{evaptime}
\frac{d\mathsf{M}}{d\tau} = -\alpha a\sigma T_H^4,
\end{eqnarray}
where $a$ is the radiation constant and $\alpha$ refers to the greybody factor.
The parameter $T_H$ represents the Hawking temperature and is given by Eq.~\eqref{TH}, $\sigma=\pi R_{\rm Sh}^2$ is the cross section as a function of the horizon radius, so one can explicitly obtain the differential equation \eqref{evaptime}, using the following value of the mass as a function of horizon radius \cite{ZhangMT,Sekhmani2023}
\begin{align}
\nonumber
\mathsf{M} = \frac{1}{6 \bar{\ell}^2 r_h}
&\biggl(
-\bar{\ell}^2 r_h^4 \left( \frac{B}{A+1} \right)^{\frac{1}{\beta +1}} \, _2F_1[\alpha, \nu; \lambda; \xi] 
\\
&\qquad \qquad\quad  
+3 \bar{\ell}^2 Q^2+3 \bar{\ell}^2 r_h^2+3 r_h^4\dfrac{}{} \biggl)\label{Mrh}.
\end{align}
Using this result of $\mathsf{M}$ in the equation~\eqref{evaptime}, one can write
\begin{eqnarray}
\int_{0}^{\tau} d\tau = -\int_{r_{\rm rem}}^{r_{\rm i}} \frac{1}{\tilde{\alpha}\sigma T_H^4} \dfrac{d\mathsf{M}}{dr_h} dr_h,
\label{evaptimer}
\end{eqnarray}
where $\tilde{\alpha} = \alpha a$, and $r_{\rm i}$ refers to the radius of the initial horizon. By integrating both sides of Eq.~\eqref{evaptimer}, the evaporation time can be found. To calculate $\tau$, an approximation to the metric of equation ~\eqref{fr} is applied, since parameter $B$ tends to zero and by setting other parameters as $\beta = 0,\, \gamma=0$ and $A>0$, MCG-AdSBH becomes the RN-AdS BH metric \cite{Sekhmani2023}.
For the RN-AdS BH metric, at $Q=0.05$ and $\bar{\ell}=20$, the evaporation time is of the order of $10^{23}/\tilde{\alpha}$.

\section{Conclusion}\label{sec9}

Motivated to reveal the MCG traces around a static and spherically charged AdS BH, we explore the dependence of the Hawking temperature and the remnant radius on the MCG-AdSBH parameters. This analysis corroborates the existence of a remnant mass and a phase transition. 
The remnant mass corresponds to the minimum mass value to which the BH can shrink, while the phase transition is associated with a maximum Hawking temperature, indicating that the corresponding heat capacity is zero at this point. 
We examine the geodesic structure around the MCG-AdSBH and investigated the far-reaching implications of this gravitational model on various astrophysical phenomena, including null trajectories, shadow silhouettes, light deflection angles, and determination of greybody bounds. 
To determine appropriate values for the model parameters, we initially derived constraints from the EHT data on the shadow radius. We identified a narrower range of parameters from observations of Sgr A*, while M87* indicated a broader range. Specifically, the  Sgr A* data provide tighter constraints on the model parameters, as they include points beyond the upper $2\sigma$ levels. 
Therefore, within a consistent parameter space for MCG-AdSBH, EHT observations do not exclude the presence of surrounding MCG at galactic centers. This study provides one of the first constraints on the modified Chaplygin dark fluid using EHT data from M87* and Sgr A*.
We then visualized how the energy emission rate varies across the parameter space and illustrated that the MCG-AdSBH parameters effectively contribute to the behavior of the greybody bounds, the emitted power, and the partial absorption cross section.
Next, we observe that in this scenario, the quasinormal modes (calculated using the eikonal approximation approach), the light deflection angle, and the black hole evaporation process exhibit sensitivity to the model parameters.
These findings are expected to contribute to our understanding  of the structure of the MCG  within the framework of modified gravity theories and their implications for BH astrophysics.

\medskip

\paragraph*{\bf{Acknowledgements}}
We would like to thank Xiao-Mei Kuang for her insightful discussions.
The research of L.M.N. and S.Z. was supported by the Q-CAYLE project,
funded by the European Union-Next Generation UE/MICIU/Plan de Recuperacion, Transformacion y Resiliencia/Junta de Castilla y Leon (PRTRC17.11), and also by RED2022-134301-T, PID2020-113406GB-I00 and PID2023-148409NB-I00, financed by MICIU/AEI/10.13039/501100011033.  The research of F.H. and H.H. was partially supported by the Long-Term Conceptual Development of a University of Hradec Králové for 2023, issued by the Ministry of Education, Youth, and Sports of the Czech Republic.

\nocite{*} % Print all references regardless of whether they were cited in the poster or not
\bibliographystyle{plain}

\end{document}